# Electronic, Wireless, and Photonic Network-on-Chip Security: Challenges and Countermeasures

**Sudeep Pasricha**, Colorado State University; **John Jose**, IIT-Guwahati; **Sujay Deb**, IIIT-Delhi

**Abstract**: Networks-on-chips (NoCs) are an integral part of emerging manycore computing chips. They play a key role in facilitating communication among processing cores and between cores and memory. To meet the aggressive performance and energy-efficiency targets of machine learning and big data applications, NoCs have been evolving to leverage emerging paradigms such as silicon photonics and wireless communication. Increasingly, these NoC fabrics are becoming susceptible to security vulnerabilities, such as from hardware trojans that can snoop, corrupt, or disrupt information transfers on NoCs. This article surveys the landscape of security challenges and countermeasures across electronic, wireless, and photonic NoCs.

## 1 Introduction

Security for emerging computing chips is becoming an increasingly important design concern. Beyond the traditional attacks from software on connected devices, attacks originating from or assisted by malicious components in hardware are becoming more common. For example, Quo Vadis Labs has reported backdoors in electronic chips that are used in weapons control systems and nuclear power plants, which can allow these chips to be compromised remotely. The well-publicized Spectre and Meltdown attacks highlight how sensitive data can be stolen from threads executing on manycore processors. It is widely acknowledged that all algorithmically secure cryptographic primitives and protocols rely on a hardware root of trust that is resilient to attacks. Clearly though, this is not the case anymore. Manycore computing chips, and the larger systems they are embedded in, are at an elevated risk for security compromises in today's world.

Networks-on-chips (NoCs) are an integral part of many-core computing chips. They play a key role in facilitating communication among on-chip cores and between cores and memory, and are key determinants of performance, energy-efficiency, and reliability at the chip-scale. Increasingly, these fabrics are becoming susceptible to security vulnerabilities, e.g., from Hardware Trojan (HTs), which are small, malicious circuits that can be inserted by untrusted third parties within a genuine electronic chip blueprint design. In NoCs, transmitted packets are often unencrypted and an HT could snoop or corrupt sensitive information which could be catastrophic. Such HTs can go untraceable during the verification and testing phase of today's global semiconductor ecosystem where it is common to outsource design automation, fabrication, and testing of integrated circuits. Introducing countermeasures to overcome these security challenges entails power and performance overheads that are not sustainable at the chip-scale. Thus, there is a need for innovative solutions to secure NoCs while keeping overheads under check. Moreover, NoCs today are evolving to consider emerging technologies such as chip-scale photonics and wireless communication. Security solutions for NoCs thus need to not only address security challenges for data transfers over electrical wires, but also over photonic and wireless channels.

This article describes security challenges in NoC design across the electrical, wireless, and photonic domains, and provides a review of promising solutions that have been proposed.

## 2 Electronic NoC Security

Over the past two decades, electronic NoCs have been the most popular communication substrate in manycore processors. Typically, NoCs connect tiles in multicore processors, where the tiles may consist of processing cores with private L1 caches or distributed L2 or L3 cache banks. NoCs are often packet switched, with each packet split into multiple smaller flow control units called flits that traverse the network. The head flit carries the control information required to forward the packet from the (upstream) source towards its (downstream) destination based on the routing algorithm, while body flits carry the data. Flits traverse one or more routers on the path to the destination. The control logic inside an NoC router consists of buffer write of incoming flits, route computation and virtual channel (VC) allocation for head flits, and switch arbitration to resolve flits with competing output port requirements. The common prefix connecting various flits of a packet consists of two fields, namely, flit type (FT) and VC identifier (VCID). FT distinguishes the type of the flit viz. head, body, or tail. VCID is computed for each head flit to indicate the buffer it should occupy upon reaching the downstream router. All other flits of the same packet inherit the same VCID. The head flit is forwarded through the NoC in an un-encrypted format due to the essential requirement of its contents needed by the intermediate routers for making rapid routing decisions. This unavoidable plain text transmission of the head flit makes it vulnerable to HT attacks.

HTs can alter the system behavior to realize attacks such as information leakage, unauthorized access, functional errors, and delay-of-service. Within NoCs, HTs can reside in the input port of a router and can access packet contents and modify selected fields of the head flit. Since NoCs support all kinds of on-chip communication packets, HTs can specifically target certain data messages, such as cache miss request packets, to create more impact. For instance, HTs that change the DID (Destination address) field of L1 miss request packets can cause significant system level performance degradation [1].

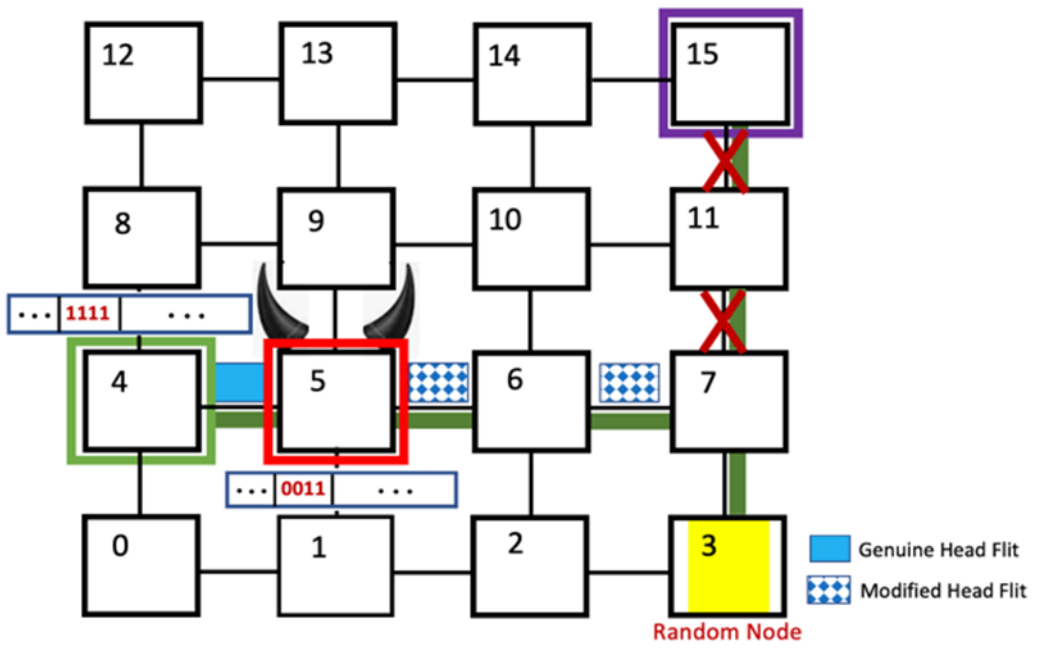

**Fig. 1: NoC with an HT in Router 5. Packet with source identifier (SID) = 4 and destination identifier (DID) = 15 is changed to DID=3 at Router 5.**

Fig. 1 shows an illustrative example of an HT in a 4x4 mesh topology based NoC. We assume that the router in tile 5 is HT infected. Consider an L1 miss request packet that is injected into router 4 and is to be routed to destination tile 15. As per XY routing, this packet passes through router 5. If the HT is active and already triggered in that router, it modifies DID to a random router number, e.g., 3. From router 6 onwards the routing is done for DID=3. Accordingly, the packet takes a south turn at router 7, reaches router 3 and subsequently gets ejected to tile 3. Here the destination tile value calculated from the address at tile 3 does not match with DID of the head flit. Hence, the packet is dropped at tile 3 without further processing. Moreover, the source of the packet (tile 4) expects a cache miss reply from tile 15 which never comes back, while tile 15 is unaware about such happenings. This leads to indefinite waiting of cache miss requests in the re-order buffer of the source tile processor, leading to application throttling.

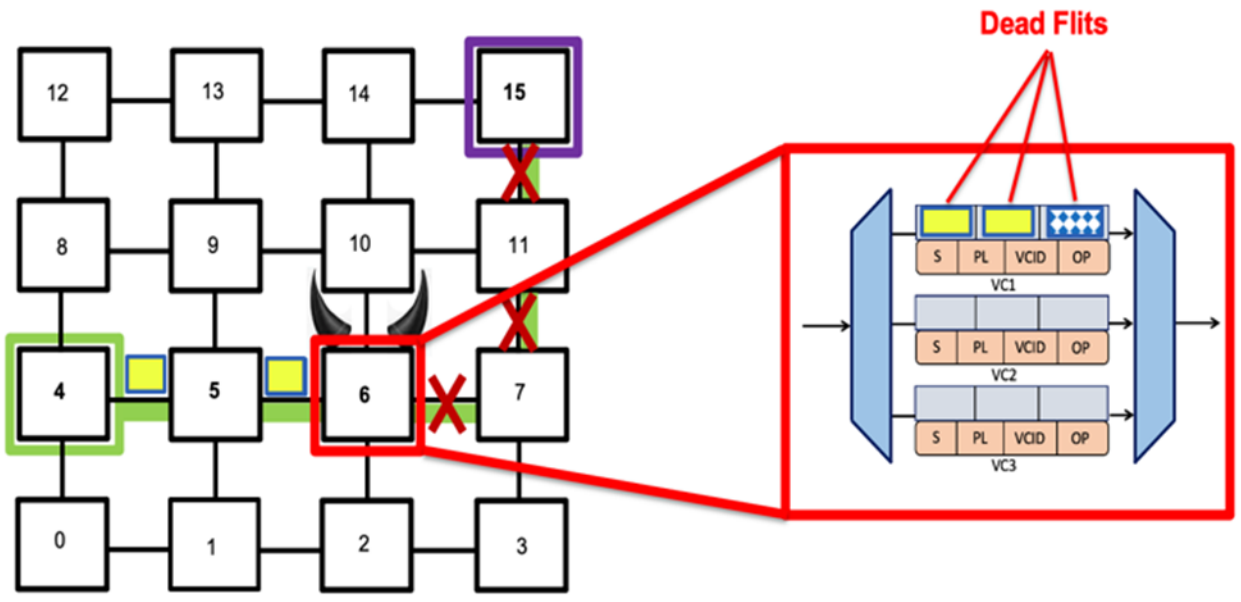

**Fig. 2: NoC with an HT in router 6 changing FT**

Fig. 2 shows an illustrative example of an HT that can manipulate the common prefix of flits [2]. Here the HT modifies a head flit into a body flit. Since the FT field indicates that it is a body flit, no route computation is done leaving the OP, VCID, and PL fields (Fig. 2) in the control buffer unset. As a direct consequence of this, the control buffer lookup for the OP field in the switch arbitration phase returns false, due to which this VC never competes for the crossbar. The flit never moves out of the buffer of the HT router and remains there forever. This eventually leads to the formation of dead flits, consuming network resources and propagating backpressure to the upstream routers. HTs can also modify routing logic. Due to wrong route computation, a packet may travel in the wrong direction and exhibit a ping-pong effect [3]. Such HTs can significantly increase packet latency. Delay HTs can induce delay while forwarding the packet to the next hop [4].

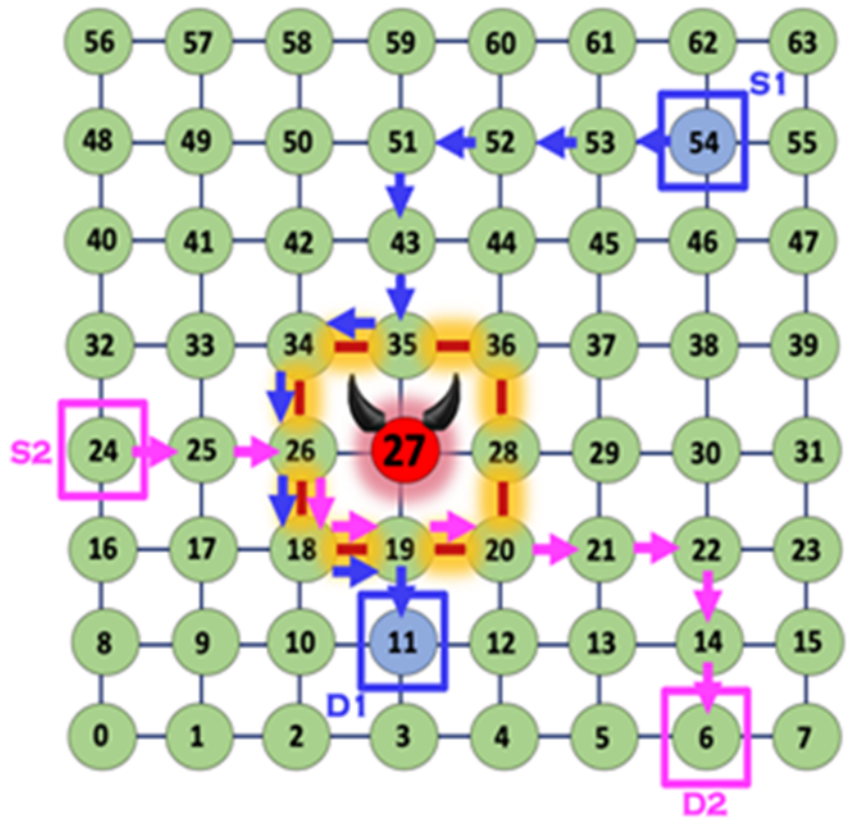

**Fig. 3: Illustration of packet rerouting in 8x8 mesh NoC**

Since any module in the NoC mesh is oblivious of the location of the HT infected node, to mitigate HT at run-time, each router needs to have a mechanism to detect the presence of HTs, to help localize the node where an HT resides, and mitigate the impact of HT behavior. HT detection and localization using caging is a promising approach [3], [4]. The HT detection can be based on hashing of message digests and lightweight encryption mechanisms. HT localization can be performed by adaptive learning on the behavior anomalies. Once the anomaly is detected by a router, special alert flits are generated to its neighboring routers. Each router, upon receiving alert flits can act either by neglecting them or keeping a watch on further alert information about potentially malicious routers. This combination of self-learning and received learning facilitates the cage formation. Once every router around an HT-infected router agrees about the presence of an HT, the caging mechanism is enabled. Mitigation mechanisms can then take care of forwarding of packets that reach the cage routers over alternative paths such that packets reach the destination bypassing the HT-infected router while taking a few extra hops. Fig. 3 illustrates the re-routing for two packets S1 and S2 after the cage is erected around HT infected router 27. When the cage is erected, packets reaching the cage-edge routers (26, 35, 19, and 28) are now re-routed. We can observe that S1 sourced at 54 and destined at 11 take a few additional hops around the cage to avoid moving through HT. Similarly, S2 also reaches its destination at 6 by going through a re-routed path.

## 3 Wireless NoC Security

In large manycore platforms, conventional electronic NoCs suffer from high power and performance overheads due to long-distance multi-hop communications. To mitigate these limitations, wireless NoC (WNoC) has been explored extensively in recent years. WNoC facilitates single-hop communication among distant intra-chip tiles through wireless direct links [5]. Multiple wireless interfaces (WIs) are optimally augmented on top of the traditional electronic NoC to build the WNoC infrastructure. These WIs are CMOS compatible and operate in the millimeter-wave (mm-wave) frequency ranges. Fig. 4 shows a WNoC topology along with the structure of the base router (BR) and the hybrid router (HR). The HR embeds the WI to the BR for enabling wireless communication.

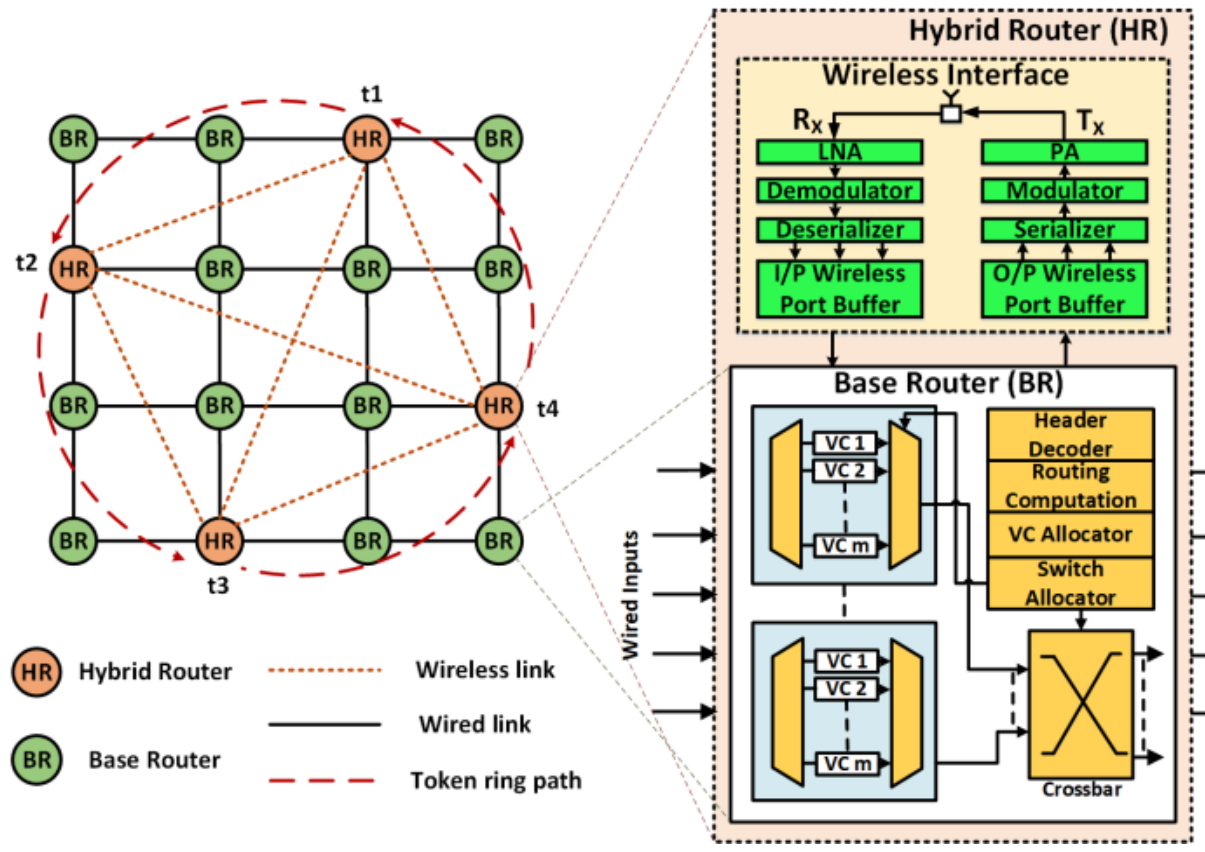

**Fig. 4: WNoC topology and hybrid router with wireless interface.**

In general, WNoCs adopt a single wireless channel to achieve low overhead implementation. Therefore, all the existing WIs need to share the same channel. A channel access control mechanism (CACM) is deployed for the fair arbitration and allocation of the wireless channel among WIs. Traditionally, a timer-based token passing CACM is used, where each WI gets the token for a fixed number of cycles in a round-robin manner as shown in Fig. 4. However, the traffic pattern and traffic density vary both spatially and temporally across the WNoC. Moreover, contemporary heterogeneous systems experience more skewed traffic. Therefore, a control mechanism that can dynamically vary the channel hold time and the token arbitration pattern based on the traffic density at the WIs is more beneficial [6]. Furthermore, in the case of a system dealing with mixed-critical data, the critical traffic density plays an important role alongside the total traffic density for varying the channel hold time and token arbitration pattern [7]. The critical load is prioritized for both WI token allocation and data transfer over the wireless channel.

Unfortunately, the CACM in a WNoC can malfunction due to a malicious HR node. An HR can be infected by the presence of an HT inside the WI or by an interfaced rogue third-party IP. A compromised CACM can change the wireless channel hold time or the token arbitration pattern in an unfair manner. This results in drastic degradation in wireless channel utilization and network throughput. For a mixed-critical data system, the impact of such vulnerabilities is more severe as the transmission of critical data gets hindered. In the case of a dynamic CACM based system, the control mechanism is mostly distributed. Therefore, it is very difficult to detect such anomalies in channel hold time and token arbitration patterns without a dedicated security framework. A malicious WI can hold the wireless channel by changing its channel access time in an unauthorized manner to create a denial-of-service (DoS) attack. Similarly, it can misguide the other WIs by changing its own source address and spoofing others by claiming unauthorized access to the wireless channel. Such attacks can be initiated by implanting a time-bomb HT as shown in Fig. 5 [8]. Whenever the register value reaches zero, the payload circuit activates the load enable signal ($LE_n$*). This loads either a new value to the access time register to create a DoS attack or a new value to the source WI register to create a spoofing attack.

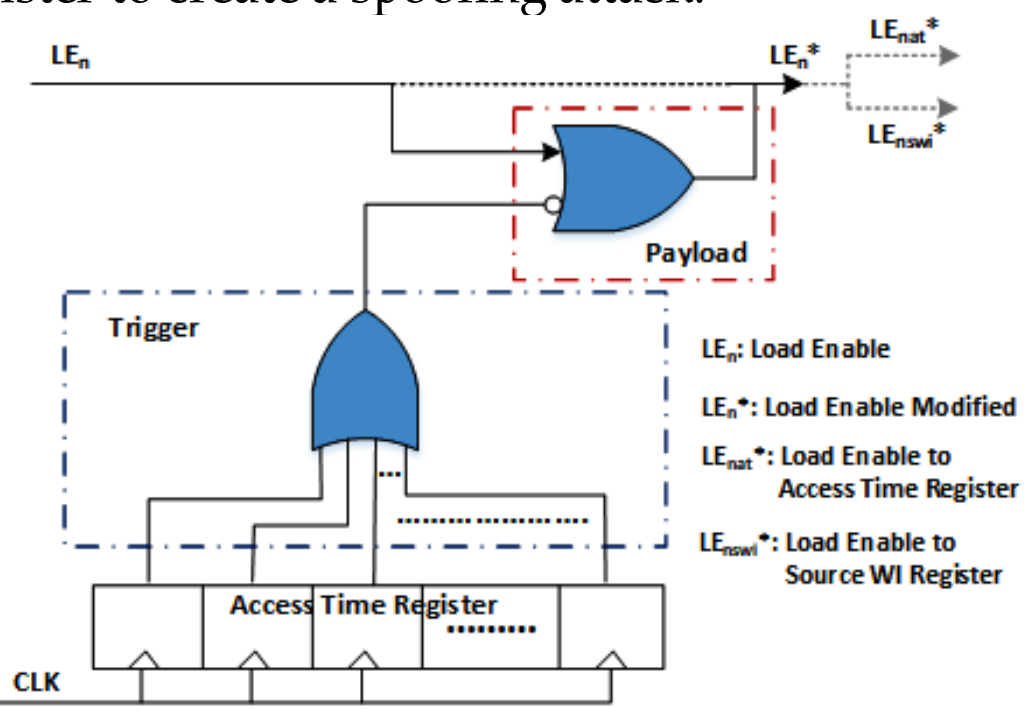

**Fig. 5: Time-bomb HT in the malicious WI [8].**

Security countermeasures to such attacks can be established based on the checking of the assigned channel access time and the actual channel access time. In a decentralized CACM, a distributed ranking-based channel access controller (DRCAC) is deployed within the WI as shown in Fig. 6(a) [8]. Initially, the information related to the traffic density and the data criticality status of all the WIs is broadcast over the wireless channel. Then, the DRCAC modules in each of the WI locally construct the WI ranking table (RT) as shown in Fig. 6(b). In case of any malicious anomalies, these RTs are used as golden references for detection and localization.

The security framework implements a flag generator and a majority voter inside the network interface (NI) of the HR nodes. The NI provides separate interfaces to the electrical router (eNI) and the WI (wNI), and is assumed to be completely secured. The flag generator module is used to create a security flag. The majority voter collects the security flags from all the WIs to make an unbiased and collective decision on the attack detection. For example, $WI_p$ being a malicious node creates a DoS attack and keeps on holding the channel beyond T1 (Fig. 6). $WI_q$, which is the next candidate in the RT, would not get the token at T1+1 cycle and would request its flag generator module to generate a security flag as shown in Fig. 6(c). The flag is sent to the other WIs over the wired NoC path. The remaining WIs would also check in their RT and generate support flags. All these flags are transferred to the HR node having $WI_p$ (the malicious one), and are provided to the corresponding majority voter module. If the majority of the flags indicate that $WI_p$ is malicious, then the voter module instructs the controller inside the NI to disconnect the WI from the rest of the network. The attack localization and countermeasure for spoofing attack works similarly.

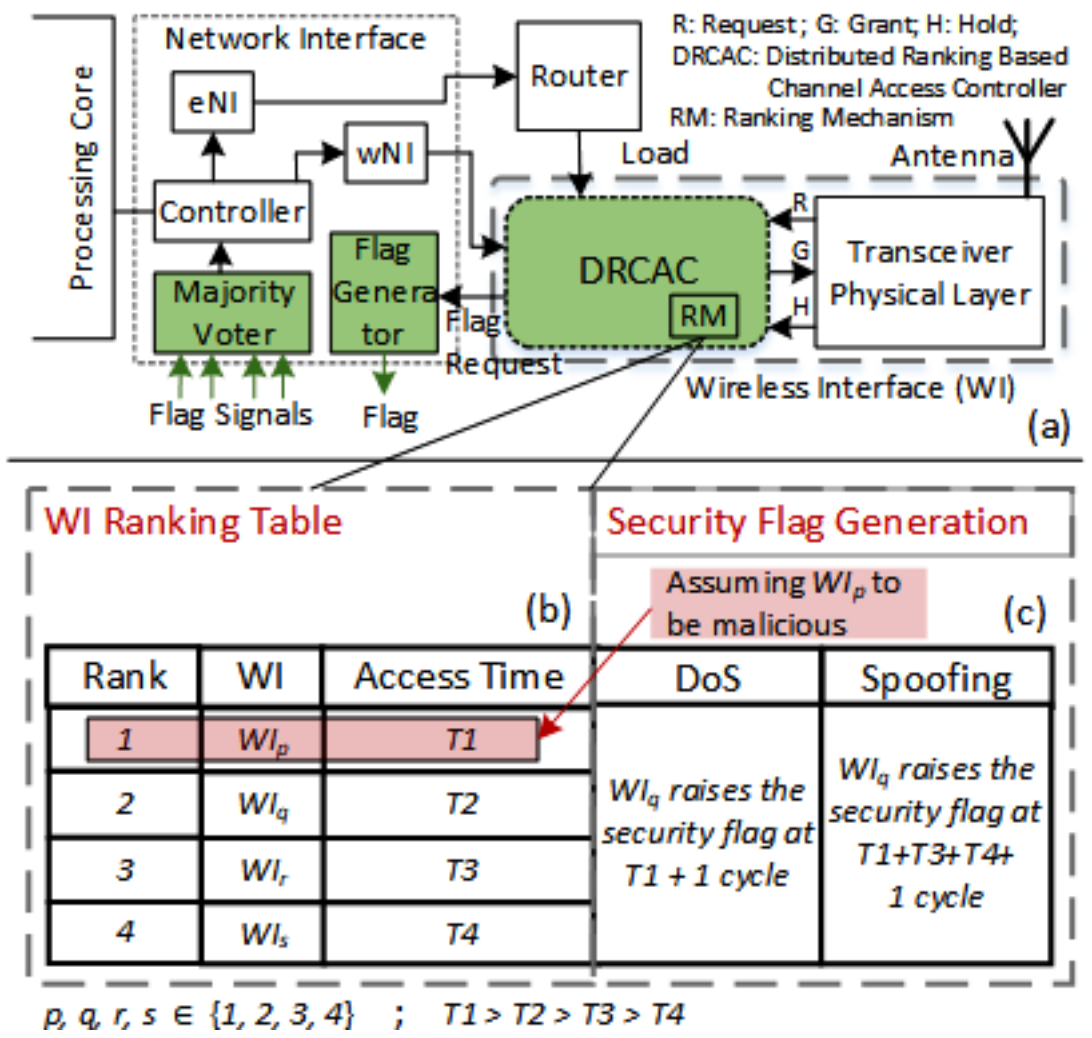

**Fig. 6: (a) Proposed ranking mechanism and security architecture, (b) Ranking table, (c) Security flag generation [8].**

The discussed threat model is a highly probable CACM attack scenario in WNoCs as a single wireless channel is shared among WIs. The presented countermeasure is lightweight, suitable for distributed CACMs, and effective in attack detection and localization in WNoCs.

## 4 Photonic NoC Security

Over the past decade, different CMOS-compatible silicon photonic devices have been developed to realize chip-scale communication in manycore computing platforms. The resulting photonic NoCs (PNoCs) provide several prolific advantages over their traditional electronic NoC counterparts, including the ability to communicate at near light speed, larger bandwidth density, and lower dynamic power dissipation [9]. PNoCs employ on-chip photonic links, each of which connects two or more gateway interfaces. A gateway interface (GI) connects the PNoC to a cluster of tiles (with cores or cache banks). Each photonic link comprises one or more photonic waveguides and each waveguide can support a large number of dense-wavelength-division-multiplexed (DWDM) wavelengths. A wavelength serves as a data signal carrier. Typically, multiple data signals are generated at a source GI in the electrical domain (as sequences of logical 1 and 0 voltage levels) which are modulated onto the multiple DWDM carrier wavelengths simultaneously, using a bank of modulator microring resonators (MRs) at the source GI. The data-modulated carrier wavelengths traverse a link to a destination GI, where an array of detector MRs filter them and drop them on photodetectors to regenerate electrical data signals. Each GI in a PNoC is able to send and receive data in the optical domain on multiple (often all) utilized carrier wavelengths. Therefore, each GI has a bank of modulator MRs (i.e., modulator bank) and a bank of detector MRs (i.e., detector bank). Each MR in a bank resonates with and operates on a specific carrier wavelength. In this manner, the excellent wavelength selectivity of MRs and DWDM capability of waveguides are utilized to enable high bandwidth parallel data transfers in PNoCs.

Unfortunately, PNoCs are also expected to incorporate the use of third-party IPs and fabrication through third party foundries in their hardware design cycle, which exposes them to security threats related to HTs. Fig. 7 shows the schematic layout of a typical PNoC, in which the gateway interfaces (GIs) are connected with each other through photonic waveguides in a serpentine topology. Each of the modulators and detectors within GIs employs control circuits to enable its active operation and redressal from process variation (PV) induced resonance shifts [10]. An adversary in the foundry can introduce HTs in these control circuits. It can also partner with software providers to introduce malicious application programs to be run on the chip hardware. As shown in Fig. 7, there can be HTs in the control circuits of multiple GIs, as well as instances of malicious programs simultaneously running on multiple cores. These HT-infected GIs can partner with malicious program instances to create security threats in the PNoC.

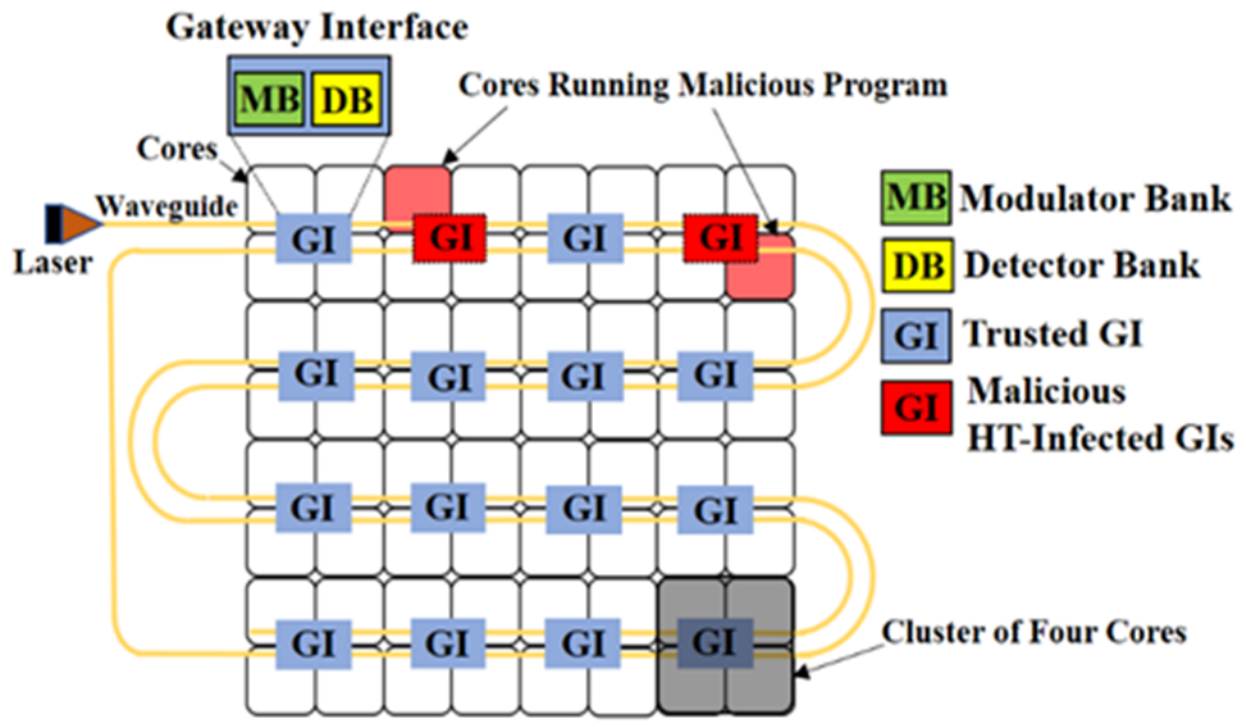

**Fig. 7: Schematic of a compromised PNoC with its processing cores running malicious software program threads and gateway interfaces (GIs) infected by Hardware Trojans (HTs).**

Fig. 8(a) and 8(b) illustrate the impact of malicious source and destination GIs on a PNoC waveguide. In Fig. 8(a), the modulator bank of source GI S1 is sending data to the detector bank of destination GI D2. When source GI S2, which is in the communication path, becomes malicious with an HT in its control logic, it can manipulate its modulator bank to modify the existing '1's in the data to '0's as part of a data integrity attack. For example, in Fig. 8(a), S1 is supposed to send '0110' to D2, but because of data corruption by malicious GI S2, '0010' is received by D2.

Let us consider another scenario for the same data communication path (i.e., from S1 to D2). When destination GI D1, which is in the communication path, becomes malicious with an HT in its control logic, the detector bank of D1 can be partially tuned to the utilized wavelength channels to snoop data, as part of a data confidentiality attack.

In the example shown in Fig. 8(b), D1 snoops '0110' from the wavelength channels that are destined to D2. The snooped data from D1 can be transferred to a malicious core to reveal sensitive information. This type of snooping attack from malicious destination GIs is hard to detect, as it does not disrupt the intended communication.

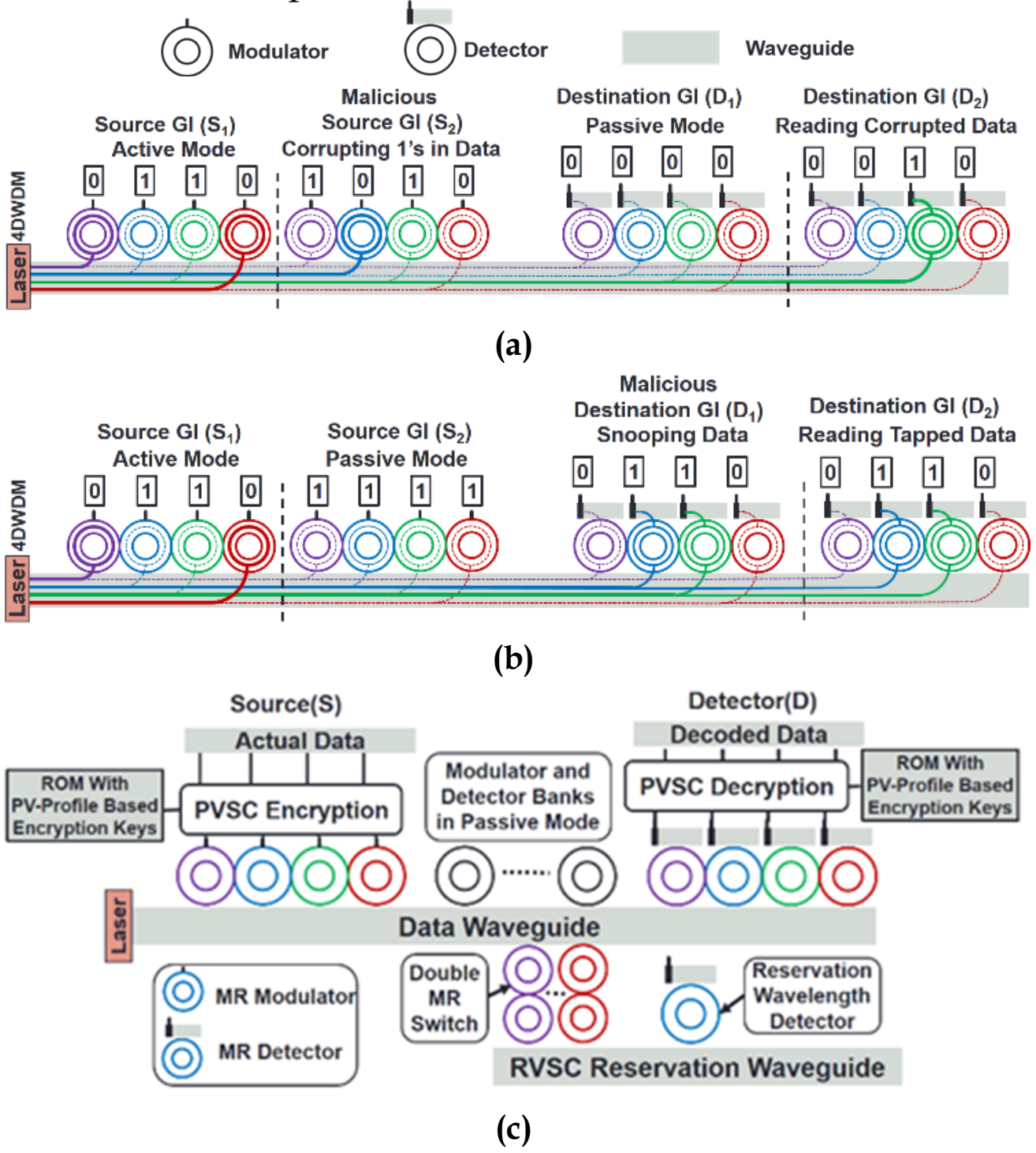

**Fig. 8: (a) Impact of malicious modulator bank, (b) Impact of malicious detector bank on data in DWDM-based photonic waveguides, (c) Overview of *SOTERIA* framework that integrates a circuit-level PV-based security enhancement scheme and an architecture-level reservation-assisted security enhancement scheme [11].**

It is also possible for an HT at a source GI (e.g., S1) to manipulate the arbitration mechanism and inject excessive spurious data on the waveguide to prevent other sources (e.g., S2) from getting access to the waveguide for legitimate transfers. Such a denial of service or data availability attack can reduce performance, which can be especially disruptive for real-time applications. Lastly, HTs at source GI can change the source information in a packet header, as part of a data authenticity attack. Such an attack can be used to orchestrate more sophisticated distributed attacks that allow stealing confidential information on the chip.

Clearly, these is a need to address the various security risks imposed by HTs that manipulate PNoC operation. Various countermeasures can be employed to protect against the different attack types.

An intuitive approach to addressing data integrity attacks is to use parity/Hamming codes in the transmitted data. If any corruption is induced on a packet by an HT, such codes at the destination GI can detect or even correct changes in data. The cost of such codes is however more significant than in electrical NoCs because the extra bits may require additional MRs and wavelengths (which in turn increases the laser power overhead).

Addressing data confidentiality attacks is a more complex problem. A promising blueprint to overcome such attacks is that proposed as part of the SOTERIA framework [11]. Fig. 8(c) gives a high-level overview of this framework. The process variation (PV) based security enhancement (PVSC) scheme uses the PV profile of the destination GI's detector MRs to generate unique keys and encrypt data before it is transmitted via the photonic waveguide. Another approach could be to enable just enough laser power (e.g., using semiconductor on-chip amplifiers [12]) for a signal so that it can arrive at the destination and be detected. If snooping occurs on this signal, it will lead to extraction of signal power by the snooping GI, leading to data corruption that can be detected using error correction codes, as discussed earlier. These schemes are sufficient to protect data from snooping GIs, if they do not know about the target destination GI. With target destination GI information, however, a snooping GI can decipher the encrypted data. Many PNoC architectures use the same waveguide to transmit both the destination GI information and actual data, making them vulnerable to data snooping attacks despite using PVSC. To further enhance security for these PNoCs, SOTERIA employs an architecture-level reservation-assisted security enhancement (RVSC) scheme that uses a secure reservation waveguide to avoid the stealing of destination GI information by snooping GIs.

To overcome authentication attacks, it is possible to extend the SOTERIA framework with the integration of a lightweight key exchange and message authentication protocol. Data availability attacks can be addressed with arbitration schemes that guarantee fairness and freedom from starvation, as well as filters at GIs that use heuristics to detect flooding scenarios and notify arbiters and other PNoC controllers to restrict traffic from GIs identified as creating denial of service conditions in a waveguide.

## 5 Conclusions and Open Challenges

Whether a NoC is implemented with electronic, wireless, or photonic building blocks, there remains a large attack surface and unique attack vectors that can be exploited by malicious actors to snoop, corrupt, and disrupt data transfers. The article covered potential NoC-centric attacks and highlighted promising approaches which can be considered as the first steps towards realizing secure and trustworthy NoCs in emerging manycore platforms. While many of the techniques are specifically tailored for a given implementation technology, there is immense potential to explore cross-fertilization of attack detection and mitigation methodologies across technologies.

There remain many open challenges in the area of NoC security. For electronic NoCs, the overhead involved in having a security framework around all NoC routers can cause performance bottlenecks. Many of the HT detection mechanisms using message authentication and cryptography schemes can affect the critical path of NoC routers thereby creating unacceptable performance overheads. So lightweight security solutions are the need of the hour. For wireless NoCs, the transceivers and their performance under process variations makes it very challenging to define appropriate countermeasures against security vulnerabilities. Also the shared nature of the wireless channels makes it inherently more prone to attacks. In photonic NoCs, fundamental silicon photonic devices are considerably sensitive to runtime thermal variations and inevitable fabrication process variations, and therefore it is imperative to analyze how attack detection and mitigation approaches are impacted by such variations. Moreover, the design and detection of HTs integrated with silicon photonic devices (such as MRs, photodetectors etc.) remains an area of open research.

## Acknowledgments

This research was supported by grants CCF-1813370 and CCF-2006788 from the National Science Foundation (NSF).

**Sudeep Pasricha** (sudeep@colostate.edu) received his Ph.D. in computer science from the University of California, Irvine in 2008. He is currently a Professor at Colorado State University. His research interests include networks-on-chip, and hardware/software co-design for energy-efficient, secure, and fault-tolerant embedded systems. He is a Senior Member of IEEE.

**John Jose** (johnjose@iitg.ac.in) is an Associate Professor in Department of Computer Science & Engineering, IIT Guwahati. He completed his Ph.D from IIT Madras in the field of computer architecture. His research interests are in computer architecture and hardware security.

**Sujay Deb** (sdeb@iiitd.ac.in) received his PhD from the Washington State University, Pullman, WA on May 2012. He is currently an Associate Professor at IIIT-Delhi. His research Interests are in Network-on-Chip (NoC), Heterogeneous System Architectures (HSA), Hardware Security etc. He is a Senior Member of IEEE.

Mail Address : 1373 Campus Delivery, Colorado State University, Fort Collins, CO 80523-1373